\documentclass[a4paper,11pt]{article}

\usepackage{stmaryrd,amssymb}
\usepackage{natbib}

\usepackage{ifthen}
\newboolean{fullversion}
\setboolean{fullversion}{true}
\newcommand{\fullversiononly}[1]{\ifthenelse{\boolean{fullversion}}{#1}{}}
\usepackage{fancyvrb}
\usepackage{listings}
\usepackage{tikz}
\usetikzlibrary{fit,shapes.geometric}
\usetikzlibrary{decorations.pathreplacing}
\usetikzlibrary{svg.path}
\usetikzlibrary{matrix}
\usetikzlibrary{calc}
\usetikzlibrary{chains, positioning}
\usetikzlibrary{arrows,automata,positioning,chains}

\newcommand{\compcert}{Comp\-Cert}
\newcommand{\coq}{Coq}
\newcommand{\ocaml}{OCaml}
\newcommand{\code}[1]{\texttt{#1}}
\newcommand{\gcc}{gcc}

\newcommand{\sem}[1]{\llbracket #1 \rrbracket}
\newcommand{\abstr}[1]{#1 ^ {\sharp}}

\newenvironment{myitemize}{\begin{list}{$\bullet$}{\leftmargin=1em \itemindent=0pt}}{\end{list}}

\title{Simple, light, yet formally verified, global common subexpression elimination and loop-invariant code motion}
\usepackage{authblk}

\author{David Monniaux}
\affil{Verimag, Univ. Grenoble Alpes, CNRS, Grenoble INP\thanks{Institute of engineering Univ. Grenoble Alpes}}
\author{Cyril Six}
\affil{Kalray S.A. \& Verimag}

\usepackage{hyperref}
\begin{document}

\maketitle

\begin{abstract}
  We present an approach for implementing a formally certified loop-invariant code motion optimization by composing an unrolling pass and a formally certified yet efficient global subexpression elimination.
  This approach is lightweight: each pass comes with a simple and independent proof of correctness.
  Experiments show the approach significantly narrows the performance gap between the {\compcert} certified compiler and state-of-the-art optimizing compilers.
  Our static analysis employs an efficient yet verified hashed set structure, resulting in fast compilation.

\end{abstract}

\noindent\emph{This is the full version, with appendices, of an article published in LCTES~2021.}

\section{Introduction}
In this article, we present an approach for obtaining a complex optimization (loop-invariant code motion), which brings a major performance boost to certain applications (including linear algebra kernels), by composition of simple optimization passes that can be proved correct using simple local arguments.
We have implemented that scheme in the {\compcert} verified compiler.%
\footnote{Our developments are available at\\\url{https://gricad-gitlab.univ-grenoble-alpes.fr/certicompil/compcert-kvx.git}}

\subsection{CompCert}
{\compcert} \cite{leroy09,Leroy-backend} is the first optimizing C
compiler with a formal proof of correctness mature enough to be used in industry \cite{airbus12,kastner:hal-01643290}; it is now available both as a research tool%
\footnote{``Vanilla'' versions: \url{https://github.com/AbsInt/CompCert}} and commercial software.%
\footnote{\url{https://www.absint.com/compcert/}}
This proof of correctness, verified by the Coq proof assistant, ensures that the behavior of the assembly code produced by the compiler matches the behavior of the source code;
in particular, {\compcert} does not have the middle-end bugs usually found in
compilers~\cite{YangCER11}.
This makes {\compcert} appealing for safety-critical embedded systems, in particular avionics \cite{airbus12,DBLP:conf/date/FrancaFLPS11}, in comparison to extant practices such as disabling almost all compiler optimizations to ease qualification.
However, {\compcert}'s moderate level of optimization compared to state-of-the-art compilers such as gcc or clang is dissuasive in other embedded but less critical contexts.%
\footnote{For instance, the seL4 formally verified microkernel was adapted to be compiled with {\compcert}, but the resulting code was too slow.}
Improving optimizations in {\compcert} is thus important for wider usage.

Most optimizations of {\compcert} operate at the level of a \emph{register transfer language} (RTL) representation, equipped with a small-step semantics:
the state of the program essentially consists of a triple $(p,r,m)$ where $p$ is a control location, $r$ is the register state and $m$ is the memory state. The semantics of loads, stores and arithmetic operations are described as mutations of parts of $r$ and $m$, and the control flow as updates to $p$.
Some transitions generate externally observable events, notably calls to external functions; the soundness of the compiler is that the sequence of these events is preserved between source and object code.

In order to prove this property, every transformation or optimization must be formally proved correct with respect to that semantics through \emph{simulations} relating steps and states before and after the transformation.
While the simulation relations for some local optimizations may be relatively simple, the ones for non-local optimizations may be quite complex.
For instance, the inlining pass, conceptually quite simple (``replace calls to certain functions by the body of these functions with appropriate renumbering of control locations and pseudo-registers'') contains approximately 560 lines of code, but its proof of correctness is approximately 2100 line long, and uses intricate simulation arguments dealing with the reorganization of memory layout due to the fusion of stack frames.
Most of such correctness arguments would typically be handwaved over in the regular compilation literature.

Long and complex proofs not only cost developer time when they are developed, they may also cause problems later when a new {\coq} version is used, a new architecture is added to {\compcert}, or when changes are made to {\compcert} internals.
There is thus a strong incentive to keep proofs short and the complexity of simulation arguments low.

An additional difficulty is that {\compcert} optimizations must be implemented in {\coq}, a strongly typed pure functional language with many onerous requirements---for instance, all recursive functions must be shown to terminate by syntactic induction, and there are no native machine integer types.
This complicates implementation and limits efficiency.
A workaround is to call {\ocaml} code from {\coq}, but in doing so one must not increase the trusted computing base, or at most by a very small and controlled amount.

\subsection{Common subexpression elimination and loop-invariant code motion}
Consider the following example (extracted from Polybench%
\footnote{\url{https://sourceforge.net/p/polybench/wiki/Home/}}):
{\small
\begin{lstlisting}
void kernel_syrk(int ni, int nj,
  DATA_TYPE alpha,  DATA_TYPE beta,
  DATA_TYPE POLYBENCH_2D(C,NI,NI,ni,ni),
  DATA_TYPE POLYBENCH_2D(A,NI,NJ,ni,nj)) {
  for (int i = 0; i < _PB_NI; i++)
    for (int j = 0; j < _PB_NI; j++)
      for (int k = 0; k < _PB_NJ; k++)
        C[i][j] += alpha*A[i][k]*A[j][k];
}
\end{lstlisting}
}

The code initially generated for the body of the innermost loop by {\compcert} computes the address of \lstinline|C[i][j]| (by sign extension, addition and multiplication through bit-shifting), reads it, does the same for \lstinline|A[i][k]| and \lstinline |A[j][k]|, performs two floating-point multiplications and one addition, then recomputes the address of \lstinline|C[i][j]| and writes to that location.

This is suboptimal. First, the address of \lstinline|C[i][j]| should be computed only once inside the loop body. Arguably, the front-end of {\compcert}, which transforms \lstinline|x += e;| where \lstinline|e| is a pure expression into \lstinline|x = x+e|, could arrange to compute the address of \lstinline|x| only once, but this is not what happens.
Instead, a \emph{local common subexpression elimination} phase, available in vanilla%
\footnote{We call ``vanilla'' the official releases of {\compcert}, as opposed to forked versions.}
{\compcert}, will notice that the second address computation, even though it is broken down into individual operators referring to different temporary variables, is the same as the first, and will reuse that address.

The address of \lstinline|C[i][j]| is a \emph{loop invariant}: it does not change along the iterations of the innermost loop.
If this address was computed just before that loop and stored into a temporary variable, then common subexpression elimination could notice that the address computation inside the loop body is identical, and thus use the value in the temporary variable instead. The computation of that address would thus be completely eliminated from the loop body.

One way to ensure that this value is computed before the innermost loop is to unroll that loop once, replacing it by:
{\small
\begin{lstlisting}
k=0;
if (k < _PB_NJ) {
  C[i][j] += alpha*A[i][k]*A[j][k];
  k++;
  for (; k < _PB_NJ; k++)
    C[i][j] += alpha*A[i][k]*A[j][k];
  }
}
\end{lstlisting}
}%
Then, the address of \lstinline|C[i][j]| is computed by the unrolled iteration and it should be possible to eliminate its computation from the loop.
What we would thus obtain is a form of \emph{loop-invariant code motion}.
      
The common subexpression elimination in vanilla {\compcert} is however too weak to notice that \lstinline|C[i][j]| in the loop is the same subexpression as \lstinline|C[i][j]| in the unfolded iteration, because it is local: it cannot propagate information across control-flow joins, including loop headers.
The reason for keeping this transformation local is that, for the analysis used, ``least upper bounds for this ordering are known to be difficult to compute efficiently'' \cite[\S7.3]{Leroy-backend}.
What is needed is a \emph{global} common subexpression elimination, capable of propagating information across control-flow joins (tests and loops).
We present here one such analysis and transformation.

Moreover, the write to \lstinline|C[i][j]| at the end of each iteration ensures that the value of \lstinline|C[i][j]| to be loaded at the beginning of each iteration (except the unrolled first one) is already available in a register, so it is possible to remove that load.
In the end, we get this AArch64 assembly code:
{\footnotesize
\begin{Verbatim}[obeytabs=true,commandchars=\\\{\},tabsize=4]
	\textcolor{blue}{sxtw}	\textcolor{blue}{x16, w4}
	\textcolor{blue}{add}		\textcolor{blue}{x5, x2, x16, lsl #13}	/* x5 := C[i] */
	\textcolor{blue}{ldr}		\textcolor{blue}{d18, [x5, w9, sxtw #3]}	/* d18 := C[i][j] */
	\textcolor{blue}{sxtw}	\textcolor{blue}{x16, w4}
	\textcolor{blue}{add}		\textcolor{blue}{x8, x3, x16, lsl #13}	/* x8 := A[i] */
	ldr		d7, [x8, #0]			/* d7 := A[i][0] */
	fmul	d3, d0, d7				/* d3 := alpha * d7 */
	\textcolor{blue}{sxtw}	\textcolor{blue}{x16, w9}
	\textcolor{blue}{add}		\textcolor{blue}{x7, x3, x16, lsl #13}	/* x7 := A[j] */
	ldr		d5, [x7, #0]			/* d5 := A[j][0] */
	fmul	d6, d3, d5				/* d6 := d3 * d5 */
	fadd	d1, d18, d6				/* d1 := d18 + d6 */
	str		d1, [x5, w9, sxtw #3]	/* C[i][j] := d1 */
	orr		w6, wzr, #1				/* k := 1 */
.L105:
	cmp		w6, w1
	b.ge	.L104
	ldr		d16, [x8, w6, sxtw #3]	/* d16 := A[i][k] */
	fmul	d4, d0, d16				/* d4 := alpha * d16 */
	ldr		d2, [x7, w6, sxtw #3]	/* d2 := A[j][k] */
	fmul	d17, d4, d2				/* d17 := d4 * d2 */
	fadd	d1, d1, d17				/* d1 := d1 + d17 */
	str		d1, [x5, w9, sxtw #3]	/* C[i][j] := d1 */
	add		w6, w6, #1				/* k := k + 1 */
	b		.L105
\end{Verbatim}
}
Until \verb+.L105+, the first iteration of the loop is unrolled, and contains computations (in blue) that later remain loop-invariant:
the addresses of \lstinline|C[i]|, \lstinline|A[j]| and \lstinline|A[i]| are computed in resp. \lstinline|x5|, \lstinline|x8| and \lstinline|x7|.
The initial value of \lstinline|C[i][j]| is also computed in \lstinline|d18| (then coerced in \lstinline|d1|).
Since these computations remain valid throughout the loop iterations, we can remove those, resulting in a loop body with fewer instructions.

\subsection{Contributions}
We propose implementing loop-invariant code motion as the composition of two simpler phases, which are proved to be correct independently of each other:
\begin{myitemize}
\item unrolling the first iteration of the loop---through a pass capable of general forms of duplication of code (Sec.~\ref{sec:duplication});
\item global subexpression elimination (Sec.~\ref{sec:CSE}).
\end{myitemize}

In this approach, as opposed to some in the compilation literature, the correctness of loop-invariant code motion does not rely on complex arguments about invariance along execution traces, but instead only on very local arguments based on lock-step simulations and dataflow analysis.

Furthermore, our global subexpression elimination eschews the efficiency issues alluded to in~\cite[\S7.3]{Leroy-backend}, yet can be quite easily proved to be correct.
This approach is of interest in itself, since it brings some performance improvement even if loop-invariant code motion is not desired.

Our global subexpression elimination internally uses a library for efficiently computing over sets of integers (Sec.~\ref{sec:hashed_sets}), also proved correct;
this is another contribution.
\fullversiononly{%
Section~\ref{sec:TCB} discusses the impact on {\compcert}'s trusted computing base of the hash-consing mechanism used for the hashed sets: basically we trust that pointer equality implies structural equality of terms.}

In Section~\ref{sec:experiments} we shall report on performance improvements in generated code, and in Section~\ref{sec:conclusion} we shall compare with other approaches and propose future extensions.

We shall now begin with an overview of the intermediate representation that we deal with in this article, and how simulations are used to prove the correctness of optimization or transformation phases over it.


\section{{\compcert}'s RTL representation}\label{sec:smallstep}
{\compcert} uses many intermediate representations, each equipped with a semantics~\cite{Leroy-backend}.
Each transformation or optimization between representations must be proved to be correct, meaning the transformed code must simulate the original with respect to observations: the sequence of calls to external functions (and assembly-level built-in functions and accesses to volatile variables) must be respected, except that undefined behavior (undefined values, trace that stops due to an error) may be replaced by arbitrary behavior.

In this article, we deal solely with the RTL (register transfer language) intermediate language, which is the one on which most optimizations already present in vanilla {\compcert} (constant propagation, local common subexpression elimination, inlining\dots), operate.

\subsection{The RTL intermediate language}
RTL views each function as a control-flow graph with a single entry point.
The nodes of the graph, labeled with positive integers, contain instructions.
Each instruction contains the identifiers of the successors of the instruction in the graph: one for most instructions, two for conditional branches (one per branch), many for jump tables, and zero for instructions that terminate the function (tail call, return).

The state of a RTL program (outside of the function call mechanism) consists of the call stack, the program counter in the current function, the values of the (pseudo) registers, and the memory.
RTL considers an unbounded number of registers, labeled by positive integers.
Each register contains a \emph{value}:
a 32-bit integer, a 64-bit integer, a 32-bit floating point, a 64-bit floating-point, a pointer, or the special ``undefined'' value.
\emph{Memory} is divided into bytes, which can be read and written as \emph{chunks} (byte, 32-bit floating-point etc.) from and to values.
  
Any analysis thus has to deal with a small variety of basic instructions and provide sound transfer functions for all of them.
We shall here focus on three of them:
\begin{description}
\item[Operation] $r_d := \textit{op}(r_1,\dots,r_n)$ where \textit{op} is an operation (which may include immediate constants), e.g. 32-bit constant, 32-bit addition or 64-bit float multiplication;
  the source operands are $r_1,\dots,r_n$ (and, for technical reasons in some cases, the memory);
  the destination is $r_d$;
  in particular there is a ``move'' operation denoted by $r_d := r_1$ that just copies data;
\item[Memory load] $r_d := \textit{chunk}[\textit{addr}(r_1,\dots,r_n)]$
  where \textit{chunk} identifies the size and type of the data being loaded (32-bit integer, 64-bit integer, 32-bit floating-point number etc.), and \textit{addr} is an addressing mode (which again may include immediate constants, such as offsets);
  the source operands are $r_1,\dots,r_n$ and the memory;
  the address used is computed from $r_1,\dots,r_n$ and the addressing mode;
  the destination is $r_d$;
  examples of addressing modes include ``add this constant to the first argument'', ``scale the second argument by the chunk size and add it to the first argument'';
\item[Memory store] $\textit{chunk}[\textit{addr}(r_1,\dots,r_n)] := r_s$
  with similar notations and meanings;
  the source operands are $r_1,\dots,r_n$;
  the destination is the memory.
\end{description}

\subsection{Lock-step simulation}
Intermediate representations in {\compcert} are connected by ``match'' relations, and code transformations must be shown to respect the ``match'' relation.
In the simplest case, the only one that we use in the optimizations that we have developed for this article, this simulation is lock-step:
``if a step $\sigma_1 \rightarrow_1 \sigma'_1$ can be taken in the first program representation, and $\sigma_1 \sim \sigma_2$, then $\sigma_2 \rightarrow_2 \sigma'_2$'' such that $\sigma_1' \sim \sigma_2'$, where $\sigma_1$ and $\sigma'_1$ are states in the first representation, $\sim$ is the ``match'' relation and $\sigma_2$ and $\sigma'_2$ are states in the second representation.

In the case of code duplication (Sec.~\ref{sec:duplication}), $\sim$ is a relation of the form ``the registers and the memory are the same, and if $p'$ is the program counter in the transformed program and $p$ the program counter in the original program, then $f(p')=p$'' where $f$ is a function mapping each control location in the transformed program to the control location in the original location from where it was copied.

In the case of common subexpression elimination (Sec.~\ref{sec:CSE}), $\sim$ is the identity between the states in the original and transformed programs (same registers, same memory, same stack) conjoined with some invariant about the values of registers (this is where the available expressions appear) and, for technical reasons, a typing invariant.

In the case of useless move cleanup (Sec.~\ref{sec:cleanup}), $\sim$ is the identity relation, except that on the register part it is extensional identity, as opposed to the default intensional identity.

\section{Code duplication}\label{sec:duplication}


Code duplication ``unrolls'' pieces of code at the RTL level, keeping instructions in the same execution order.
We rely on the \textit{a posteriori verification} technique to prove it correct: some untrusted {\ocaml} function transforms the code, then a formally proven verifier in {\coq} either accepts or rejects the transformed code.
In this work, we use this pass for unfolding the first iteration of innermost loops and for ``rotating'' loops.

\subsection{Unfolding the first iteration of innermost loops}

First, we identify innermost loops using standard algorithms on control-flow graphs.
In \compcert, \texttt{for} and \texttt{while} loops are generated as follows: first the computation of the parameters of the condition expression, then
a conditional branch to either exit the loop, or go onto the next instructions, which we name \textit{loop body}.
Finally, the last instruction of the \textit{loop body} has a backedge to the start of the loop.

\begin{figure}[tb]
  \begin{tikzpicture}[->,>=stealth',shorten >=1pt,auto,node distance= 0.35cm and 0.6cm, semithick,
    nstyle/.style={draw, align=left},
    blue/.style={color=blue},
    red/.style={color=red!50},
    emap/.style={red, dashed},
    ndup/.style={nstyle, blue},
    edup/.style={blue}]
    \node [nstyle] (A) {Pre-computing\\condition};
    \node [nstyle] (B) [below= of A] {Loop condition};
    \node [nstyle] (C) [below= of B] {Loop body};
    \node [nstyle] (D) [right= of C] {Exit};

    \path 
      (A) edge (B)
      (B) edge (C)
          edge (D)
      (C) edge [bend left=90] (A);

    \node [nstyle] (A2) [right= 1.7cm of A] {Pre-computing\\condition};
    \node [nstyle] (B2) [below= of A2] {Loop condition};
    \node [nstyle] (C2) [below= of B2] {Loop body};
    \node [nstyle] (D2) [right= of C2] {Exit};

    \path 
      (A2) edge (B2)
      (B2) edge (C2)
           edge (D2)
      (C2) edge [bend left=90] (A2);

    \node [ndup] (C') [above= of A2] {Loop body};
    \node [ndup] (B') [above= of C'] {Loop condition};
    \node [ndup] (A') [above= of B'] {Pre-computing\\condition};

    \path
      (A') edge [edup] (B')
      (B') edge [edup] (C')
           edge [edup, bend left=40] (D2)
      (C') edge [edup] (A2);

    \path
      (A'.west) edge [emap] (A.east)
      (A2.west) edge [emap] (A.east)
      (B'.west) edge [emap] (B.east)
      (B2.west) edge [emap] (B.east)
      (C'.west) edge [emap] (C.east)
      (C2.west) edge [emap, bend right] (C.east)
      (D2.west) edge [emap, bend left] (D.east);
  \end{tikzpicture}
  \caption{The first iteration of the original code (on the left) is unrolled,
  resulting in the code on the right.
  The duplicated instructions are in blue.
  The \textit{reverse mapping} $f$ is in red.}
  \label{fig:duplication}
\end{figure}
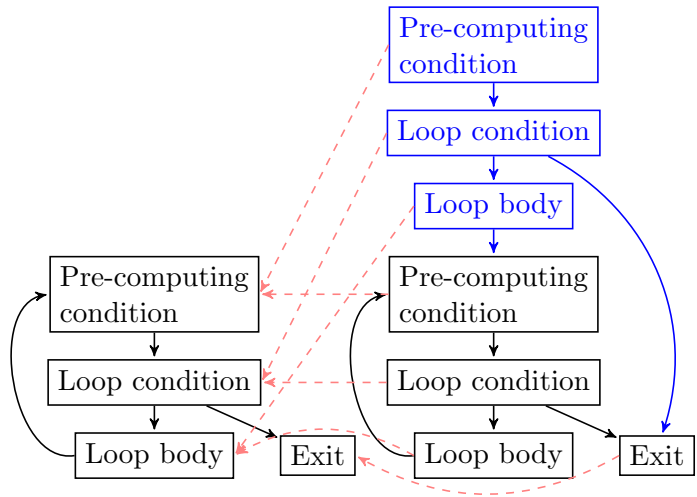

Unfolding the first iteration consists of duplicating that code, and setting the successor of the duplicated \textit{loop body} to the actual loop, instead of a backedge (Fig.~\ref{fig:duplication}).
It amounts to replacing \lstinline|while(c) {b}| with
\lstinline|if(c) { b; while(c) {b}}|.

Since this part is handled in untrusted {\ocaml} code, we do not need to prove any property about these algorithms.%
\footnote{A bug in these algorithms might result in a worst performance, but never to incorrectness, since that would be caught by the certified verifier.}
However, to guide the verifier in knowing which nodes were duplicated, the oracle exhibits a \textit{reverse mapping} $f$: for every control location $p'$ of the transformed code, $f(p')$ is the origin of the copy in the original code.

\subsection{Formally checked verifier}

Once the untrusted code returns a transformed code (and a \textit{reverse mapping} $f$), our formally verified checker ensures that this new code is indeed faithful to the original code, in terms of order of execution of the duplicated instructions.
We do so by comparing each instruction $p'$ of the transformed code against the supposedly original instruction $f(p')$ from the original code.
This check only succeeds if both instructions are the same, modulo the following property: "Each pair of successors $(s', s)$ of $(p', f(p'))$ must verify $f(s') = s$".
For example, if $p'$ is a conditional branch, then $f(p')$ must be a conditional branch with the same parameters except the successors: if we denote $(s_{true}, s_{false})$ the successors of $f(p')$ and $(s_{true}', s_{false}')$ the successors of $p'$, then we must have $f(s_{true}') = s_{true}$ and $f(s_{false}') = s_{false}$.

The verifier is then certified with a lock-step simulation, where the \textit{match} relation ($\sim$) holds if two states $\sigma$ $\sigma'$ have all of their terms equal but their program counters $p$ and $p'$, which must instead verify the relation $f(p') = p$.

This allows verifying any transformation that duplicates instructions and changes some of the successors to point to duplicated instructions instead of original code.
In particular, it can be used for other transformations such as tail-duplication (for a superblock scheduling); we do not cover these here.

\subsection{Loop rotation}
Another optimization that can be performed and verified by our duplication scheme is \emph{loop rotation}.
Straightforwardly compiling a loop \lstinline|while(condition) { body }| produces one conditional branch and one unconditional branch:
{\small
\begin{lstlisting}
head: if (! condition) goto exit;
      body; goto head; exit:
\end{lstlisting}}
\noindent It is often slightly better to ``rotate'' the loop into:
{\small
\begin{lstlisting}
if (condition) {do {body} while(condition);}
\end{lstlisting}}
\noindent This way, the evaluation of the condition may be merged and scheduled into the body, and there is only one single (conditional) branch.

\section{Common subexpression elimination}\label{sec:CSE}
The main difficulty in implementing optimizations in {\compcert} is to keep the complexity of the proofs low.
One way to do so is to split them into several phases, each with a clear specification.
Our CSE3 common subexpression elimination is thus implemented in four steps:
\begin{enumerate}
\item an untrusted analysis collects inductive invariants and stores them in an efficient format (hash-consed sets);
\item a verified checker checks that these invariants are truly inductive;
\item a verified code transformation phase replaces redundant computations by ``move'' operations, assuming the above invariants are correct;
\item a verified code transformation replaces moves from one variable to itself by ``no operation'' instructions;
\item a verified code transformation removes dead code.
\end{enumerate}
For simplicity of implementation, most of the code of the first two steps is shared.

\subsection{Untrusted static analysis}
\subsubsection{Abstract domain and semantics}
Our abstract domain collects equalities of the general form $r_d = \textit{rhs}(r_1,\dots,\allowbreak r_n)$ where $r_d,r_1,\dots,r_n$ are pseudo-registers, and with two subtypes: either \textit{operations} or \textit{loads}. The notations and meanings are similar to RTL:
\begin{description}
\item[Operation] $r_d = \textit{op}(r_1,\dots,r_n)$;
\item[Memory load] $r_d = \textit{chunk}[\textit{addr}(r_1,\dots,r_n)]$.
\end{description}
The semantics of an equality is the set of pairs (registers, memory) that match the equality: both sides of the equality evaluate to the same value.
For instance, for a load $r_d = \textit{chunk}[\textit{addr}(r_1,\dots,r_n)]$, the set of pairs (registers, memory) is such that evaluating the addressing mode over the values of the registers $r_1,\dots,r_n$, then loading the chunk at that address in the memory, yields a value equal to the value of register~$r_d$.%
\footnote{Loads from an invalid address are defined to yield the special value ``undefined''. This is necessary in order to accommodate the ``non trapping'' or ``dismissible'' load instructions found in certain architectures, which return a default value instead of trapping on incorrect memory references. Such instructions are useful to anticipate loads before conditional branches.}

The analysis attaches a set of such equalities to each control location in the function under analysis (the analysis is performed independently for each function).
We invoke CompCert's implementation of \citet{10.1145/512927.512945}'s algorithm (in the forward direction), which solves a system of dataflow equations in a semi-lattice.
This algorithm may equivalently be understood as a simple solver for monotone fixed-point equations in the abstract interpretation style in a finite-height lattice~\cite{Cousot78} (just use the dual of the data-flow lattice).

The semantics $\sem{S}$ of a set of equalities $S$ is the intersection of the semantics of the equalities in the set; that is, it is the set of pairs (registers, memory) that match all the equalities in~$S$.

For efficiency reasons, each equality occurring within the analysis of a function%
\footnote{Our analysis is intraprocedural, as often for common subexpression elimination.}
is uniquely identified by a positive integer.
A set of equalities is thus represented as a set of positive integers;
equality between two sets of equalities is thus equality between two sets of positive integers.
After a control-flow merge, only the equalities present in all incoming sets are conserved, thus the control-flow merge or least upper bound operation amounts to set intersection.
Since equality tests and least upper bounds occur very frequently, we opted for hash-consed integer sets (Sec.~\ref{sec:hashed_sets}).


\subsubsection{Transfer functions}
\paragraph{Assignments}
The transfer function for the operation $r_d := \textit{op}(r_1,\dots,r_n)$ for $r_d \notin \{ r_1,\dots,r_n \}$ first discards all equalities involving $r_d$, then adds an equality $r_d = \textit{op}(r_1,\dots,\allowbreak r_n)$.
For instance, for the operation $r_1 := r_2*5$ (\textit{op} is then the unary operation $x \mapsto 5x$; recall the operator may contain immediate constants) we generate the equality $r_1 = r_2*5$.

In order to quickly find the positive integer associated to this equality, a hash table of all equalities created so far is maintained, along with a counter for the next available equality identifier: if an equality does not already exist in the table, it is added to it and associated with the current value of the counter, which is incremented.
To be able to discard all equalities involving $r_d$ in one set difference operation, for all register $r$ the set of all extant equalities involving $r$ is maintained and updated as new equalities are created.

The transfer function for a load $r_d := \textit{chunk}[\textit{addr}(r_1,\dots,\allowbreak r_n)]$ where $r_d \notin \{ r_1,\dots,r_n \}$ proceeds similarly.
For an operation or load such that  $r_d \in \{ r_1,\dots,r_n \}$ we just discard all equalities involving~$r_d$.

\paragraph{Memory store}
A sound but coarse transfer function for a ``store'' $\textit{chunk}[\textit{addr}(r_1,\dots,r_n)] := r_s$ operation is to discard all equalities involving memory, most notably the loads.%
\footnote{For technical semantic reasons that we shall not discuss here, some arithmetic operations are also considered by {\compcert} to depend on memory, thus equations involving them must be discarded on a memory write.}
Again, in order to do so efficiently by set difference, the set of all extant equalities involving memory is maintained and updated as new equalities are created.

A refinement of this approach applies an alias analysis: the intersection of the set of present equalities and the set of equalities involving memory is computed; an equality of the form $r'_d = \textit{chunk}[\textit{addr'}(r'_1,\dots,r'_n)]$ is then discarded only if the alias analysis cannot prove that $\textit{chunk}[\textit{addr}(r_1,\allowbreak\dots,\allowbreak r_n)]$ and $\textit{chunk}[\textit{addr'}(r'_1,\dots,r'_n)]$ cannot overlap.
Currently this analysis is very simple: it states that memory references within two different global symbols do not overlap, and that memory blocks at two non-overlapping index ranges relative to the same base pointer (e.g. array accesses at different constant offsets, accesses to different fields within the same structure) do not overlap.

A further refinement is to consider that a store $\textit{chunk}[\allowbreak\textit{addr}(r_1,\dots,r_n)] := r_s$ induces an equality $r_s = \textit{chunk}[\textit{addr}\allowbreak(r_1,\dots,r_n)]$.
There are a few subtleties here.
First, this is only true if the chunk is 32-bit or 64-bit, or, with 8-bit and 16-bit integer writes, the value being read is not the original 32-bit integer that was in $r_s$, but rather its low-order bit truncation.
In addition, since {\compcert}'s RTL is untyped, there could be semantic mismatches if an ill-typed operation was executed (e.g. $r_s$ contains a floating-point value but \textit{chunk} is integer).
We thus run {\compcert}'s typing analysis first and verify that the type of the \textit{chunk} matches the type computed for~$r_s$.%
\footnote{If {\compcert}'s type analysis fails, due to some variable being used to store values of different types, our optimization phase fails.
  This is consistent with {\compcert}'s register allocation failing if the program is ill-typed.
  {\compcert}'s RTL generation phase always produces correctly typed programs, and all optimization phases should maintain this typing property.
  This is an example among others of an invariant that {\compcert} expects to be maintained, and that is checked dynamically.}

\paragraph{Function calls}
The user selects, by a command-line option, to model function calls by forgetting all relations, or just those involving memory. The latter will try to conserve values in registers across calls and thus increase register pressure, which may be detrimental. A possibility would be for an oracle estimating register pressure to make that choice.

\paragraph{Move-forwarding}
We do not apply the transfer functions described above directly.
We also first \emph{forward} their operands: each $r_i$ in the right-hand side is possibly replaced by $r'_i$ so that there is a ``move'' equation $r_i = r'_i$ in the current set (these are for instance generated from assignments $r_i := r'_i$).
To quickly obtain these ``move'' equations, we take the intersection of the current set of valid equations with the set of identifiers of ``move'' equations with $r$ on the left-hand side.
This means that we must maintain for all $r$ the set of all identifiers of such equations.

\paragraph{Recognition of already computed expressions}
When processing an assignment $r_d := \textit{op}(r_1,\dots,r_n)$  (respectively, load $r_d := \textit{chunk}[\textit{addr}(r_1,\dots,r_n)]$),
such that $\textit{op}$ is not a ``move'',
we first look for an equation $r'_d = \textit{op}(r_1,\dots,r_n)$ (respectively, $r'_d = \textit{chunk}[\textit{addr}(r_1,\dots,r_n)]$) in the current set.
If one exists, then in addition to the $r_d = \textit{op}(r'_1,\dots,r'_n)$
equation, we also add the equation $r_d = r'_d$,
which may be useful later for move-forwarding (this can be disabled through a command-line option).

Again, in order to find suitable equation identifiers, we intersect the current set of valid equations with the set of identifiers of equations with the suitable right-hand side.
To do so, we maintain a hash table mapping each possible right-hand side to the set of equations in which it appears.

\subsubsection{Tables to maintain}\label{sec:analysis_tables}
Our analysis is untrusted and implemented in {\ocaml}, therefore we have access to all of {\ocaml} features, including efficient imperative hash tables.

The analysis maintains:
\begin{myitemize}
\item a hash table mapping each equation to its identifier, a positive integer, with automatic allocation and assignment to a fresh identifier if the equation is not yet in the table;
\item conversely, a \emph{catalog} map from identifiers to the associated equation;
\item a hash table for mapping each equation right hand side to the set of identifiers of equations having this right hand side;
\item for each $r$, the set of all identifiers of equations involving~$r$;
\item for each $r$, the set of all identifiers of ``move'' equations of the form $r = r_1$;
\item the set of all identifiers of equations involving memory.
\end{myitemize}

While, for efficiency, all these tables are created empty and updated dynamically as new equations are discovered, their contents can all be recomputed from the catalog.
This property will be used for verified analysis.

\subsubsection{Final result}
The final result of the untrusted analysis is composed of a few read-only data structures:
\begin{myitemize}
\item the catalog of equations
\item the table mapping equations to identifiers
\item the table mapping right-hand sides to sets of identifiers
\item the inductive invariants, as a map from program locations to sets of identifiers
\end{myitemize}
No hypothesis (logical axiom) will be made about the contents of these structures in the verified parts of the analysis and the transformation.%
\footnote{The read-only hash tables are exported to Coq as their ``find'' operation: functions mapping a key to an optional value.
  There is an implicit logical assumption that this ``find'' operation behaves as a pure function.%
\fullversiononly{See Section~\ref{sec:TCB} for a discussion of this.}}

\subsection{Inductiveness check}
From the catalog of equations produced by the static analysis, we recompute various tables in a formally verified manner, using {\coq} code:
\begin{myitemize}
\item for each $r$, the set of all identifiers of equations present in the catalog involving~$r$;
\item for each $r$, the set of all identifiers of ``move'' equations of the form $r = r_1$ present in the catalog;
\item the set of all identifiers of equations involving memory present in the catalog.
\end{myitemize}
By ``formally verified'', we mean we prove theorems stating that the sets that we compute contain the sets describe above; e.g., the ``the set of all identifiers of equations involving~$r$'' that we compute truly contains the set of all identifiers of equations present in the catalog involving~$r$.
We need these properties to prove the soundness of the transfer functions.

We then check that the invariants produced by the static analysis are truly inductive, using transfer functions implemented in {\coq}.%
\footnote{For ease of implementation, the transfer functions used in the verified inductiveness check and those used in the untrusted static analysis are the same {\coq} code.}
We prove soundness theorems about these functions, in the usual abstract interpretation fashion: if the program can take a step $\sigma \rightarrow \sigma'$ through an instruction $I$, and $\sigma \in \sem{S}$, and $\abstr{I}$ is the abstract transfer function associated with instruction $I$, then $\sigma' \in \sem{\abstr{I}(S)}$.

The inductiveness check just boils down to checking (again, using verified {\coq} code):
\begin{myitemize}
\item that the ``top'' element of the abstract lattice is associated to the function entrypoint (any values in the registers, any values in the memory, no known relation between them)
\item that if there is an instruction edge from control location $p$ to control location $p'$, labeled with instruction $I$, and $p$ is labeled with $S_p$ and $p'$ is labeled with $S_{p'}$, then $\abstr{I}(S_p) \sqsubset S_{p'}$ where $\sqsubset$ is the ordering in the lattice.
\end{myitemize}
Through standard interpretation formalism, this entails that at any control location $p$, labeled with $S_p$, any state reachable at this location belongs to $\sem{S_p}$: the $S_p$ form a system of inductive invariants.

\subsection{Code transformation}
Our code transformation preserves the structure of the function;
it replaces some instructions (operations and loads) $r_d := \textit{rhs}$ with ``move'' operations if the result of the instruction already exists in one current register.

In order to do so, the transformation applies ``move forwarding'', as in the static analysis, then
computes the intersection of the set of equations whose identifiers appear in the invariant associated to the control location of the instruction and the set of equations whose right hand side match the right hand side of the instruction.
If one equation $r'_d = \textit{rhs}$ is found, then the instruction is replaced by a move $r_d := r'_d$.

As in {\compcert}'s original CSE, some operations (e.g. loading immediate constants) are deemed ``trivial'', meaning they cost so little that it is not worth replacing them by moves of available expressions. These operations are not replaced.

The correctness proof is a basic lock-step simulation between deterministic programs: one step in the source program maps to one step in the transformed program, with register and memory states matching exactly.
This correctness proof uses the fact that the $S_p$ are invariants of the program.
For instance, the reason why it is legal to replace $r := a+b$ by $r := x$ is that, using these invariants, we know that at this point in the program, the sum of the values of registers $a$ and $b$ is always equal to the value of register~$x$.

When several operations are replaced by moves, some of these moves may themselves become redundant. For instance, a memory access \lstinline|t[a*i+b]| may be compiled into
{\small%
\begin{lstlisting}
ai = a*i; aib = ai+b;
addr=t+aib<<3; r=*addr,
\end{lstlisting}
}%
\noindent so a second identical access is compiled into
{\small%
\begin{lstlisting}
ai2 = a*i; aib2 = ai2+b;
addr2=t+aib<<3; r2=*addr2
\end{lstlisting}
}%
\noindent Our analysis replaces the operations in this second access by
{\small%
\begin{lstlisting}
ai2 = ai; aib2 = aib; addr2=addr; r2=r
\end{lstlisting}
}%
Variables \lstinline|ai2|, \lstinline|aib2|, \lstinline|addr2| are ``dead'' and are discarded along with the assignment to them by a later cleanup phase.

\subsection{Cleanup phases}\label{sec:cleanup}
The map from registers to values in the state is viewed \emph{intensionally} (two maps are equal if and only if their internal structure is equal).
This means that writing $m[x]$ into a map $m$ at position $x$ returns a map that is not in general equal to $m$.
This is not a problem for our proofs, except for one step: replacing assignments $x := x$ (which may be generated by common subexpression elimination) by ``no operation''.
This needs an \emph{extensional} view, where two maps are considered to be equal if and only if they are equal at every position.
This extensional view can be defined as an equivalence relation over maps, compatible with the map operations.

One approach would have been to define the simulation relation for common subexpression elimination using this equivalence relation instead of map identity, but this would have tended to make all proofs heavier even though we need extensionality only for generating ``no operation'' instead of $x := x$ assignments. Instead, we opted for a separate phase that replaces these assignment with ``no operation'', proved correct using a lockstep simulation relation based on this equivalence relation.

We then use {\compcert}'s dead code elimination to remove useless ``moves'' produced by~CSE3.

\section{Hash-consed integer sets}\label{sec:hashed_sets}
{\compcert} provides a library (\code{Maps.PTree}) of trees with nodes indexed by positive integers, defining partial maps from the positive integers to an arbitrary type~$A$.
In {\compcert}, a positive integer is uniquely defined by the sequence of its binary digits starting from the least significant, and ending by a~1.
This sequence of digits is used as a path from the root of a binary tree: $1$ corresponds to the root, $2$ to its first child, $3$ to its second child, etc.
A tree thus consists either in an ``empty'' leaf, or in a node pointing to a ``0'' subtree (if the next digit in the sequence is $0$), to a ``1'' subtree (if the next digit in the sequence is $1$) and containing an optional element from~$A$.
If $A$ is chosen to be the ``unit'' type, then these trees implement sets of integers (as sets of keys associated to nodes with the optional element from the unit type): a present optional value indicates ``true'', an absent value ``false''.

There are however two limitations to this approach:
\begin{myitemize}
\item the representation is not unique: there are infinitely many representations of the empty set (all trees whose nodes contain no optional element);
\item many operations (equality test, inclusion test, union, intersection) are trivial if their operands are equal, but there is no fast way for recognizing this case.
\end{myitemize}
To overcome both, we use an approach similar to the ``smart constructor'' approach advocated by \citet{DBLP:journals/jar/BraibantJM14} for implementing verified reduced ordered binary decision diagrams in Coq.

The first limitation is overcome by adding the constraint that the tree should be \emph{reduced}:
the tree should not include any node pointing to two ``empty'' leaves and containing the Boolean ``false''.
We design all functions producing trees so that they automatically reduce the nodes they create, and prove theorems of the form ``if the trees passed to this function are reduced, then its output is reduced''.
Finally, we wrap the trees so that only reduced trees are available externally: a set of positive integers is represented as a dependent pair, the first element is a tree $t$, the second a proof that $t$ is reduced.%
\footnote{Such pairs can be conveniently used in lieu of the trees themselves, two sets being semantically equal if and only if the associated pairs are equal, without the need of adding the axiom of proof irrelevance. Indeed, reducedness is a decidable property $P$, so a proof that $t$ is reduced is just a proof that $P(t)=\code{true}$.
  The Boolean type obviously has decidable equality, and it is a theorem (\code{Coq.Logic.Eqdep\_dec.eq\_proofs\_unicity\_on}) that if $a$ belongs to a type with decidable equality, there is a unique proof that $a=a$
  (in other words, Streicher's axiom K is actually a theorem on types with decidable equality).}

The second limitation is overcome by \emph{hash-consing} the nodes, ensuring that there are never, at a given moment, two copies of the same tree residing at different memory locations inside the OCaml program extracted from Coq.
This is achieved by telling Coq's extraction mechanism to replace the normal constructor (and also, for technical reasons, the match operation) over the tree data type with a constructor that looks up a global hash table for a node isomorphic to the one being created, and returns the extant isomorphic node if it exists, otherwise adding the newly created node to the table.
The hash table is weak, meaning that OCaml's garbage collector is allowed to remove elements from it if they become otherwise unreachable (a normal hash table would prevent useless nodes from being collected).
This is the only addition we make to {\compcert}'s trusted computing base (TCB): we trust this hash table. \fullversiononly{More information about {\compcert}'s TCB in \autoref{sec:TCB}.}

We could reduce further the TCB by making our hashed set library less generally usable by not providing a constant-time equality test (physical pointer equality).
The property that we really use about hash-consing is that pointer equality implies structural equality, which is not an issue.
The other property that hash-consing guarantees, that structural equality implies pointer equality, involves the correctness of the hashing mechanism and the fact that we never create objects outside of that mechanism, a bigger addition to the TCB;
but we do not actually need that property: in our usage, it is equality of objects that allow optimizations, not inequality.

The isomorphism test for hash-consing is shallow: hash-consed nodes are isomorphic if and only if their contain the same Boolean, their left subtrees point to the same location, their right subtrees point to the same location.
Each node also contains a hidden ``unique identifier'' field, containing a 64-bit number allocated at node creation (a global counter is incremented at each creation), so as to make hashing shallow as well:
the hash value of a node is a hash of the triple composed of the Boolean in the node and the unique identifiers at the roots of its children.
In order for nodes to be collected as garbage if they become useless, we use OCaml's weak hash tables: pointers from the hash table to memory blocks do not cause these blocks to be considered in use.

The Coq development follows these lines:
first, the tree structure is defined along with its semantics: given a positive integer $i$ and a tree $t$, whether $i$ belongs to the set defined by $t$.
A structural equality test (tree isomorphism) is defined and proved to be correct; then, all set operations (inclusion, union, intersection, subtraction) are defined, using the structural equality test to trigger shortcuts (e.g. $a \cap b = a$ if $a = b$).
This is inefficient as a pure Coq implementation, since the linear-time structural equality test is triggered at every recursion step of the operations;
but during extraction, this structural equality is replaced by an extremely fast call to OCaml pointer equality (\code{==}).

Again along the lines of {\compcert} extant \code{Maps.PTree} module, a ``contents'' function is provided, producing a list whose contents (defined using Coq's classical {\code{In} predicate) is provably identical to the set contents.
A ``fold'' operator is provided, shown to be provably equivalent to taking the contents and running the classical left fold operation on lists.

\section{Experiments}\label{sec:experiments}

\newcommand{\perftable}[1]{%
  \begin{tabular}{l|rr|rr|rr}
    \multicolumn{1}{c|}{Benchmark} &
    \multicolumn{4}{c|}{\compcert} &
    \multicolumn{2}{c}{\gcc}\\
    &
    \multicolumn{2}{c|}{no unroll} &
    \multicolumn{2}{c|}{unroll+rotate} & \\
    & CSE3 & SSA
    & CSE3 & SSA
    & -O1 & -O2\\         
    \hline
    \input{benchmark_data/measures_#1}
  \end{tabular}
}

\begin{table}
\begin{center}\small
  \begin{tabular}{l|rr|rr|rr}
    \multicolumn{1}{c|}{Benchmark} &
    \multicolumn{4}{c|}{\compcert} &
    \multicolumn{2}{c}{\gcc}\\
    &
    \multicolumn{2}{c|}{no unroll} &
    \multicolumn{2}{c|}{unroll+rotate} & \\
    & CSE3 & SSA
    & CSE3 & SSA
    & -O1 & -O2\\         
    \hline
    glpk & 0.99 & 1.00 & 0.96 & 0.97 & 1.01 & 0.97\\
picosat & 0.98 & 1.03 & 0.99 & 0.99 & 0.73 & 0.72\\
genann4 & 0.98 & 1.00 & 0.89 & 0.91 & 0.92 & 0.70\\
jpeg-6b & 1.05 & 1.01 & 0.98 & 0.95 & 1.02 & 0.81\\
ocaml & 1.03 & 1.00 & 1.03 & 1.03 & 0.94 & 0.86\\
\hline
correlation & 0.99 & 1.00 & 0.99 & 1.00 & 0.91 & 0.90\\
covariance & 0.99 & 1.00 & 0.99 & 1.00 & 0.91 & 0.89\\
2mm & 0.99 & 0.99 & 0.99 & 0.99 & 0.98 & 0.97\\
3mm & 0.99 & 0.99 & 0.99 & 0.99 & 0.98 & 0.98\\
atax & 0.93 & 0.85 & 0.71 & 0.69 & 0.71 & 0.69\\
bicg & 1.00 & 1.00 & 0.80 & 0.80 & 0.83 & 0.80\\
cholesky & 0.89 & 0.89 & 0.62 & 0.62 & 0.68 & 0.62\\
doitgen & 0.92 & 0.94 & 0.84 & 0.84 & 0.71 & 0.71\\
gemm & 0.99 & 0.99 & 0.98 & 0.98 & 0.97 & 0.97\\
gemver & 0.99 & 1.00 & 0.97 & 0.97 & 0.94 & 0.93\\
gesummv & 0.97 & 1.00 & 0.83 & 0.83 & 0.86 & 0.83\\
mvt & 0.99 & 1.00 & 0.98 & 0.98 & 0.95 & 0.94\\
symm & 1.00 & 1.00 & 0.76 & 0.76 & 0.98 & 0.98\\
syr2k & 0.98 & 0.98 & 0.89 & 0.80 & 0.83 & 0.80\\
syrk & 1.00 & 1.00 & 0.88 & 0.70 & 0.75 & 0.70\\
trisolv & 1.00 & 1.00 & 0.70 & 0.70 & 0.85 & 0.71\\
trmm & 0.96 & 0.96 & 0.66 & 0.66 & 0.70 & 0.66\\
durbin & 0.88 & 0.87 & 0.89 & 0.88 & 1.00 & 0.87\\
dynprog & 0.79 & 0.77 & 0.62 & 0.63 & 0.61 & 0.33\\
gramschmidt & 0.76 & 0.76 & 0.76 & 0.75 & 0.98 & 0.75\\
lu & 0.87 & 0.88 & 0.67 & 0.67 & 0.68 & 0.67\\
ludcmp & 0.95 & 0.97 & 0.98 & 0.98 & 0.86 & 0.83\\
floyd-warshall & 0.77 & 0.73 & 0.66 & 0.52 & 0.55 & 0.55\\
reg\_detect & 0.93 & 0.90 & 0.38 & 0.35 & 0.34 & 0.33\\
adi & 0.98 & 0.98 & 0.95 & 0.93 & 0.89 & 0.87\\
fdtd-2d & 0.97 & 0.97 & 0.85 & 0.86 & 0.78 & 0.77\\
jacobi-1d-imper & 0.99 & 0.99 & 0.85 & 0.85 & 0.80 & 0.76\\
jacobi-2d-imper & 1.00 & 0.93 & 1.00 & 0.99 & 0.72 & 0.71\\
seidel-2d & 0.99 & 0.99 & 0.97 & 0.99 & 0.84 & 0.86\\

  \end{tabular}

\end{center} 
\caption{Performance on AArch64 Cortex-A53.  All numbers give the time (in cycles) relative to the baseline: our version of {\compcert} without SSA or CSE3.\\Top: application benchmarks, below: Polybench.\\
  SSA performs global value numbering (GVN) and sparse conditional constant propagation (SCCP). GVN has about the same effect as CSE3.
  ``Unroll'' means that the first iteration of each innermost loop under a threshold size is unrolled, allowing, together with CSE3 or GVN,
  loop-invariant code motion.}
\label{tab:perf_aarch64}  
\end{table}

We evaluated our common subexpression elimination and loop-invariant code motion schemes on several architectures.
On all of them,
the combination of unrolling the first loop iteration and global subexpression elimination dramatically increases the speed of certain benchmarks.

\subsection{Benchmark configurations}
\paragraph{Hardware}
We ran benchmarks on ARM Cortex~A53 (AArch64) inside a Raspberry Pi~3 running Ubuntu GNU/Linux 18.04.5~LTS. This dual-issue, in-order  core was chosen because it is similar to other in-order ARM cores used in embedded systems; also it is used as little core in ``big.LITTLE'' settings; gcc~8.3.0-2.
In addition, we experimented on x86-64 (Xeon Gold 6138), Risc-V (``Rocket'') and Kalray KV3 cores\fullversiononly{ (see \ref{sec:extra_perf})}.
In each case, we tie the process to one core of the machine, and we measure clock cycles using hardware counters.

\paragraph{Benchmarks}
We used the Polybench/C~3.2 benchmark suite%
\footnote{\url{http://web.cse.ohio-state.edu/~pouchet.2/software/polybench/}} as well as a few fuller-scale applications:
\begin{myitemize}
\item The GNU Linear Programming Toolkit (GLPK) v4.65%
\footnote{\url{https://www.gnu.org/software/glpk/}}
  solving one if its benchmarks (``prod''),
\item Libjpeg-6b,%
\footnote{\url{http://libjpeg.sourceforge.net/}}
  the reference JPEG implementation, compressing one of its test images;
\item Picosat v965,%
\footnote{\url{http://fmv.jku.at/picosat/}}
  a SAT-solver, solving a sudoku problem encoded into CNF-SAT;
\item OCaml%
\footnote{\url{https://ocaml.org/}}
  runtime system v4.07.1,
  running the bytecode of a quicksort implementation on a sample list;
\item Genann, a neural network library.%
  \footnote{\url{https://github.com/codeplea/genann}}
\end{myitemize}

For Polybench, we use the standard dataset except on Risc-V (small dataset), due to instability of the platform, and on KV3 (mini dataset) due to memory limitations in our evaluation board setup.

\subsection{Performance results}
Performance on Cortex-A53 is shown in Table.~\ref{tab:perf_aarch64}\fullversiononly{ (\emph{other architectures in \autoref{sec:extra_perf}})}.
Our loop invariant code motion and common subexpression elimination scheme improves performance by 10\% to 20\% on average,%
\footnote{Geometric means of the ratios across all benchmarks}
depending on the architecture.
CSE3 alone does not procure much of that speed gain, as most of the eliminations it can do alone are already done by {\compcert}'s extant CSE;
adding unrolling and thereby apply loop invariant code motion gains 12\% speed over CSE3 alone on Cortex-A53.
On Cortex-A53 and KV3, our improved {\compcert} produces code only 10\% slower than \verb|gcc -O2|.

\begin{center}
\begin{tabular}{l|rrr|rrr}
CPU & \multicolumn{6}{c}{Differences in cycles spent (\%) compared to}\\
& \multicolumn{3}{c|}{no CSE3, no unroll} &
      \multicolumn{3}{c}{\texttt{gcc -O2}} \\
& avg & min & max & avg & min & max\\
Cortex-A53  & -16 & -63 & +3 & +10 & -23 & +87\\
Rocket      & -10 & -43 & +1 & +29 & 0 & +184 \\
Xeon        & -21 & -56 & +4 & +21 & -3 & +189 \\
KV3         & -11 & -32 & +3 & +8 & -13 & +88 \\
\end{tabular}
\end{center}


\begin{figure}
\includegraphics[width=\columnwidth]{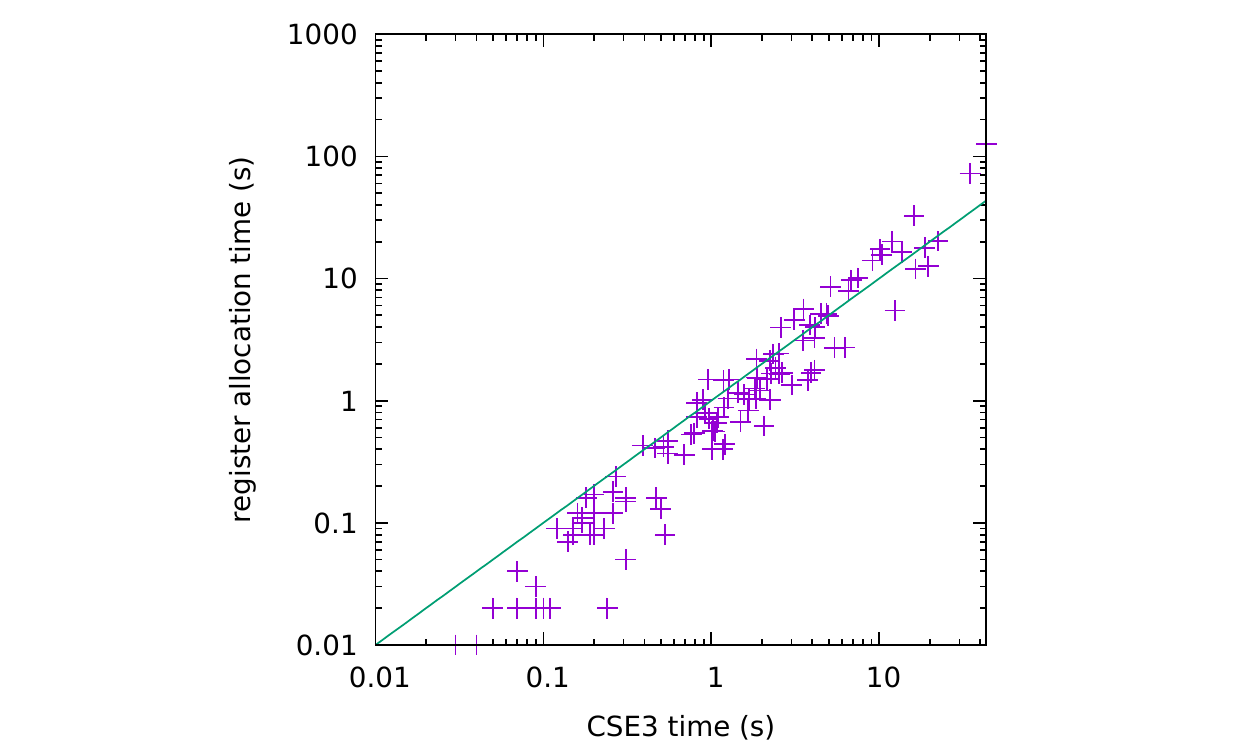}\\[-1em]
  \caption{CSE3 vs register allocation compilation time}
  \label{fig:CSE3_vs_regalloc}
\end{figure}

\begin{figure}
  \begin{center}
    \includegraphics[width=\columnwidth]{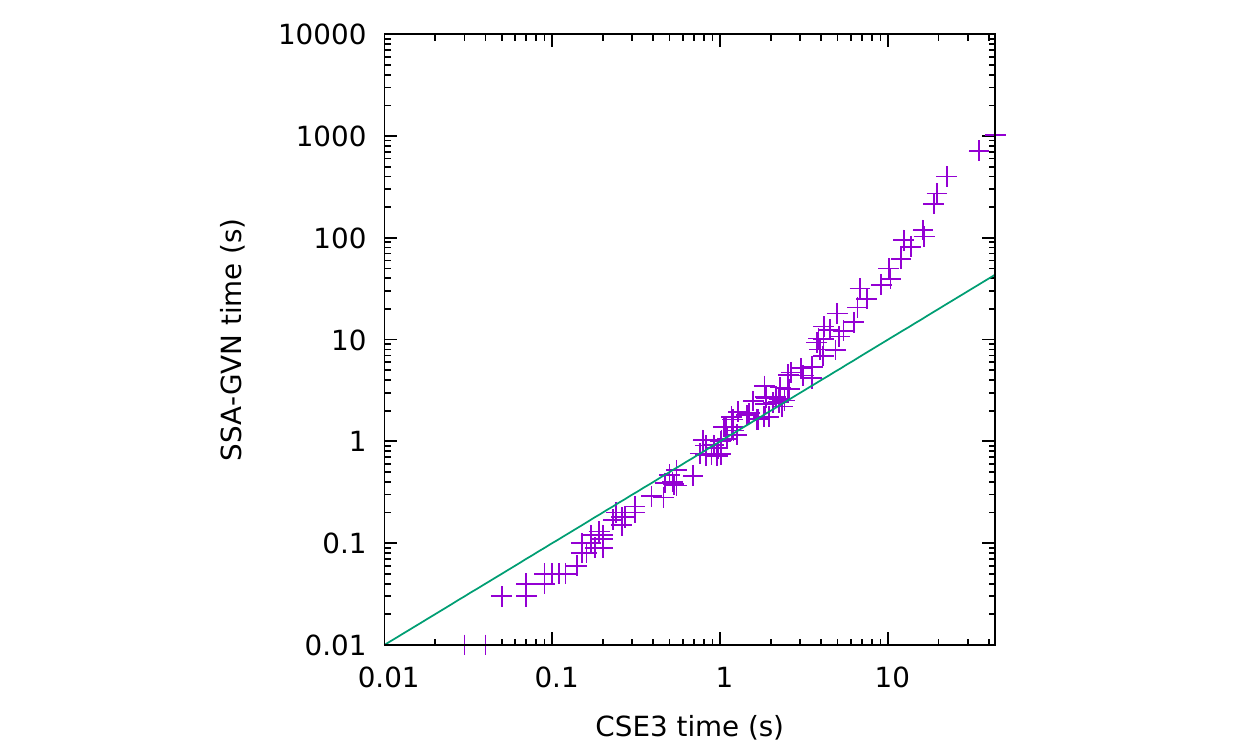}
  \end{center}
  \caption{CSE3 vs SSA-GVN compilation time}
  \label{fig:CSE3_vs_SSA-GVN}
\end{figure}

We also verified that our approach with hashing is way faster than one without hashing, and that in any case the running time of our transformation is dominated by that of register allocation on larger functions (\autoref{fig:CSE3_vs_regalloc})\fullversiononly{; see also \autoref{sec:compilation_speed}}.

\subsection{Comparison with {\gcc}}
It is difficult to identify reasons for relative slowness on larger ``application'' benchmarks, because there are so many possible optimizations that may affect the result (inlining strategy).
On the smaller Polybench benchmarks, with tight loops, we have identified the following useful optimizations not currently implemented by {\compcert}.

\paragraph{Strength reduction of address computations on loop indices}
Polybench contains many accesses to arrays. An access \lstinline|a[i][j]|, where \lstinline|i| and \lstinline|j| are 32-bit integers, to a bidimensional array \lstinline|a|, induces an address computation $a+(\mathit{ext}(i)\times N+\mathit{ext}(j)) \times S$ where $a$ is the base address of \lstinline|a|, $N$ is the number of columns (in which $j$ ranges), $S$ is the size in bytes of one array cell, including padding, and $\mathit{ext}$ is the sign extension function from 32-bit integers to 64-bit integers (assuming 64-bit pointers).
{\compcert} issues all these operations into RTL for every access, and it is up to RTL optimizations to identify that some of these operations are redundant.

Multiplication is typically much slower than addition. \emph{Strength reduction} replaces a multiplication $i \times N$, where $i$ is a loop index incremented by $K$ at every iteration, by an extra variable $i_K$ incremented by $KN$ at every iteration.

\paragraph{Integer size promotion}
Consider the loop
where \lstinline|i| and \lstinline|n| are 32-bit integers not overwritten within the loop body: \lstinline|for(int i=0; i<n; i++) { .. }|.
Equivalently, one could convert \lstinline|n| to a 64-bit integer before the loop, then use \lstinline|i| as a 64-bit integer.
This would avoid sign extension instructions within the loop body.

\paragraph{Advanced loop optimizations}
With the same loop as above, the trip count of the loop is known to be \lstinline|n| and this allows many optimizations, including \emph{software pipelining} (starting some operations for the next loop iteration, such as fetching data, within the current one), using hardware loops on architectures supporting them (KV3), etc.




\section{Related work, prospects, and conclusion}\label{sec:conclusion}
There are very few formally verified compilers. Early (1980-1990s) prototypes of verified compilers tended not to include optimizations. The two major current verified compilers are {\compcert} and CakeML. CakeML does not feature common subexpression elimination.
Two less mature projects of verified compilers, Velus\footnote{\url{https://velus.inria.fr/}}
and CertiCoq,\footnote{\url{https://certicoq.org/}}
use {\compcert} as a backend; our optimizations benefit them.
Velus \cite{Velus_POPL20} compiles a subset of the Lustre data-flow synchronous language, similar to industrial languages such as Scade or Simulink meant for implementing control laws in embedded systems.

Formally verified compilation is still a challenge. Classical optimizations, available in mainstream compilers, may be surprisingly difficult to prove correct.
\citet{DBLP:conf/pldi/TristanL09} proposed a system for lazy code motion inside {\compcert}. This system was not made available, and in particular was never integrated into {\compcert}, in particular because of high cost on large functions.%
\footnote{see Leroy's answer \url{https://github.com/AbsInt/CompCert/issues/274}}
Tristan's thesis states that their available expression analysis, used in lazy code motion, takes cubic time~\cite[\S 5.4.4.]{DBLP:phd/hal/Tristan09}.
It is difficult for us to compare our work to \citeauthor{DBLP:conf/pldi/TristanL09}'s since their publications give a high level view, missing important details, and their implementation is not available.
Their proof is much bigger than ours despite the algorithmics being less efficient.

In modern compilers, the strongest forms of common subexpression elimination (global value numbering, etc.) and of code motion are often implemented on some \emph{single static assignment} (SSA) form~\cite{SSAbook}.
A ``middle-end'' based on conversion to SSA, optimization, the conversion from SSA, was implemented into CompCert \cite{DBLP:journals/toplas/BartheDP14,DBLP:conf/cc/DemangePS15}, and was recently ported to current versions of {\compcert}.%
\footnote{\url{https://gitlab.inria.fr/compcertssa/compcertssa}}
This is certainly a more general approach than ours, but also much heavier.
The SSA middle-end, including global value numbering, comprises about 53,000 lines of code, whereas our common subexpression elimination is only 2,700 line long, to which must be added 1,500 lines for the hashed set library (which is completely independent of the rest of {\compcert} and thus immune to changes in semantics, architectures etc.).
Their correctness proofs are harder and involve non-trivial invariants about control-flow graphs, dominance relations, etc.
Furthermore, their system is significantly slower compared to ours (\autoref{fig:CSE3_vs_SSA-GVN}); further investigation is needed to establish why.

Instead of unrolling the first iteration of the loop, we investigated (and even implemented) injecting a copy of the possibly loop-invariant statements as dead code, writing to fresh variables, before the loop entry, and then using common subexpression elimination to replace instructions in the loop body by moves from these fresh variables; then dead code elimination will erase the injected statements that are not actually used.
This approach however suffers from several shortcomings:
\begin{myitemize}
\item one cannot move memory loads out of loops, because they may trap if the memory location is incorrect;%
  \footnote{Unless the instruction set has dismissible loads, that is, loads that return a default value instead of trapping.
    The only architecture with such instructions that is supported by {\compcert} (not in ``vanilla'') is the Kalray~KV3, in certain modes of operation.}
\item the same for trapping arithmetic instructions (e.g., division on some architectures, because of division by zero);
\item one cannot remove loads from memory of values that have just been written to by the preceding iteration.
\end{myitemize}

One may object that unrolling the first iteration of a loop may increase the code size needlessly, even when there is little loop-invariant code
(object files inflate by an average of 5\% on AArch64 with our settings).
Future work will include an untrusted check for loop-invariant values before unrolling the first iteration and/or a pass that would roll back needlessly unrolled iterations (this is the reverse of code duplication and thus can be verified as easily).

An alternative to verified compilation is \emph{translation validation}: the program is compiled with a conventional compiler, then the object and source code are compared by a tool.
\citet{DBLP:conf/pldi/SewellMK13} successfully applied this approach to a 10000-line microkernel (seL4).
The approach must be tuned according to the compiler used and uses heuristics that may break with some optimizations.
The fact that this approach was not ported to programs other than seL4 seems to indicate that it is limited in applicability and/or that significant efforts are needed for each new program to be compiled.

\bibliographystyle{plainnat}
\bibliography{CSE3_PLDI2021}

\appendix

\section{Compilation speed}\label{sec:compilation_speed}

\begin{figure}
  \begin{center}
    \includegraphics[width=\columnwidth]{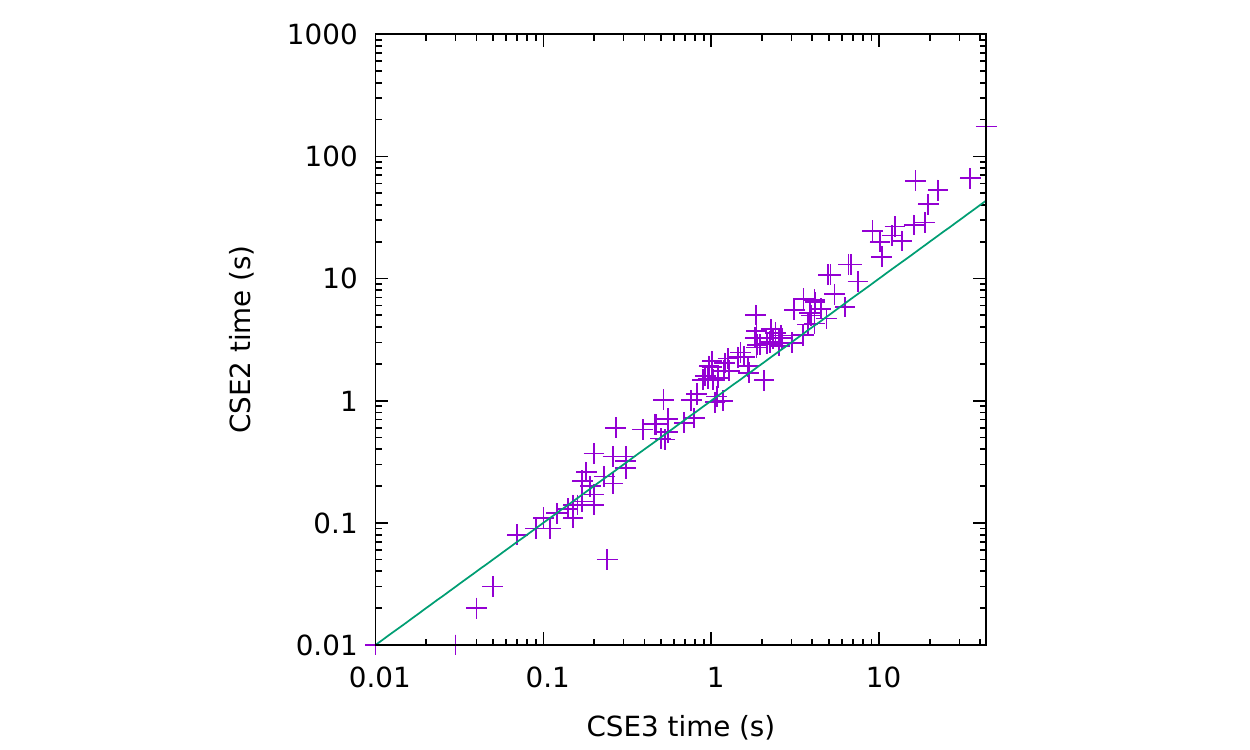}
  \end{center}
  \caption{CSE3 vs CSE2 compilation time}
  \label{fig:CSE3_vs_CSE2}
\end{figure}

We timed various {\compcert} phases for x86-64 on a family of programs generated by Yarpgen%
\footnote{\url{https://github.com/intel/yarpgen}},
a tool for testing compilers.
CSE2, a simpler version of the same analysis, without hashed sets, is slower (Fig.~\ref{fig:CSE3_vs_CSE2}); this justifies the use of hashed sets.
Register allocation is also slower than CSE3 (Fig.~\ref{fig:CSE3_vs_regalloc}).

\section{Necessary precautions}\label{sec:precautions}
Comparing the performance of machine code generated by different compilers is fraught with difficulties.
Here are some precautions we had to take.

\paragraph{Scheduling}
The Cortex-A53, KV3 and Rocket processors execute instructions \emph{in-order}.%
\footnote{The KV3 uses very large instruction words (VLIW), meaning that the compiler can schedule several instructions at the same clock cycle. The Cortex-A53 can issue two consecutive instructions at the same clock cycle if they are compatible.
  However, they will not reorder instructions.}
If the operands of an instruction are unavailable because they have not yet been written out by preceding instructions, the processor will stall
(an out-of-order processor such as the Xeon Gold may start executing following instructions).
Seemingly unimportant changes in the generated code (e.g. out-of-SSA writing out instructions in another order) may thus, especially in tight loops, result in notable differences in execution times.

Optimizing compilers, including {\gcc}, \emph{schedule} instructions to minimize stalls.
The Kalray~KV3 port of {\compcert} schedules instructions inside basic blocks, after register allocation~\cite{six:hal-02185883}, but such post-pass scheduling is not available for other architectures.
It would be unfair to compare performance between {\gcc} with scheduling and {\compcert} without.

Because of these two reasons, we run our experiments with a pre-pass (before register allocation) instruction scheduler. It formally checks that the instructions are properly reordered in a manner similar to~\cite{six:hal-02185883}; it will be covered in another publication.
This experimental scheduler was developed without access to microarchitectural documentation; improvements may thus be expected in the future.


\paragraph{Contracted floating-point expressions}
{\compcert}, at least the formally verified parts of it, is not allowed to modify the semantics of program constructs except by refinement: it can allow executions with undefined behaviors to proceed past that them, and can replace undefined values by arbitrary values---but it cannot replace a well-defined value by another.
The C standard, however, sometimes allows altering semantics.
This is in particular the case with \emph{contracted expressions} \cite[\S 6.5, \S F.7]{C18}, i.e., replacing $a \times b+c$ by a \emph{fused multiply add}:
\begin{quote}
  A floating expression may be contracted, that is, evaluated as though it were a single operation, thereby omitting rounding errors implied by the source code and the expression evaluation method.
\end{quote}
In order to keep performance results comparable, we disable contracted expressions in {\gcc} using \verb|-ffp-contract=off|.

\paragraph{x86-64: bad fit for CISC}
{\compcert} was designed for RISC processors with separate instructions for accessing memory.
In contrast, x86 and x86-64 have instructions that access memory and perform arithmetic, for instance load a value from memory and add it to a register.
{\compcert} will not use these instructions and instead go through a temporary register.
It is unclear how much this reduces the performance of the code produced by {\compcert}. 

\section{Performance measurements on other platforms}
\label{sec:extra_perf}
\subsection{Kalray KV3}
Kalray KV3, a manycore processor, with in-order, very large instruction (VLIW) cores; gcc 7.5.0.

\begin{center}
  \begin{tabular}{l|rr|rr|rr}
    \multicolumn{1}{c|}{Benchmark} &
    \multicolumn{4}{c|}{\compcert} &
    \multicolumn{2}{c}{\gcc}\\
    &
    \multicolumn{2}{c|}{no unroll} &
    \multicolumn{2}{c|}{unroll+rotate} & \\
    & CSE3 & SSA
    & CSE3 & SSA
    & -O1 & -O2\\         
    \hline
    glpk & 0.99 & 1.00 & 0.98 & 0.98 & 1.03 & 0.94\\
picosat & 0.99 & 1.01 & 0.99 & 0.99 & 1.03 & 0.96\\
genann4 & 0.98 & 0.99 & 0.86 & 0.89 & 1.07 & 0.67\\
jpeg-6b & 1.01 & 1.01 & 0.97 & 0.97 & 1.21 & 0.91\\
ocaml & 1.00 & 1.00 & 1.01 & 1.01 & 1.10 & 1.05\\
\hline
correlation & 1.00 & 1.00 & 1.00 & 1.00 & 0.61 & 0.53\\
covariance & 0.95 & 0.95 & 0.94 & 0.99 & 1.18 & 0.78\\
2mm & 0.92 & 0.92 & 0.86 & 0.89 & 1.16 & 0.80\\
3mm & 0.94 & 0.94 & 0.89 & 0.89 & 1.20 & 0.76\\
atax & 0.88 & 0.88 & 0.83 & 0.86 & 1.15 & 0.95\\
bicg & 0.97 & 0.97 & 0.94 & 0.94 & 1.14 & 0.94\\
cholesky & 0.90 & 0.89 & 0.74 & 0.74 & 1.10 & 0.71\\
doitgen & 0.94 & 0.95 & 0.95 & 0.94 & 1.04 & 0.68\\
gemm & 1.00 & 0.89 & 0.89 & 0.89 & 1.21 & 0.83\\
gemver & 1.00 & 0.97 & 0.92 & 0.92 & 1.25 & 0.97\\
gesummv & 1.00 & 1.00 & 0.97 & 0.97 & 1.14 & 0.94\\
mvt & 0.94 & 0.97 & 0.94 & 0.92 & 1.22 & 0.83\\
symm & 1.00 & 1.00 & 0.91 & 0.92 & 1.03 & 0.82\\
syr2k & 1.03 & 1.00 & 1.03 & 1.00 & 1.09 & 0.91\\
syrk & 1.00 & 1.00 & 0.85 & 0.85 & 1.11 & 0.80\\
trisolv & 0.92 & 0.93 & 0.81 & 0.82 & 1.09 & 0.80\\
trmm & 1.00 & 1.00 & 0.76 & 0.76 & 0.99 & 0.73\\
durbin & 0.99 & 0.99 & 0.88 & 0.87 & 1.27 & 0.82\\
dynprog & 1.00 & 0.95 & 0.85 & 0.81 & 1.42 & 0.75\\
gramschmidt & 0.99 & 0.94 & 0.94 & 0.94 & 1.12 & 0.94\\
lu & 1.00 & 1.00 & 0.74 & 0.74 & 1.00 & 0.76\\
ludcmp & 0.95 & 0.94 & 0.77 & 0.76 & 0.96 & 0.66\\
floyd-warshall & 1.00 & 1.00 & 1.00 & 1.00 & 0.86 & 0.76\\
reg\_detect & 0.91 & 0.92 & 0.68 & 0.72 & 0.93 & 0.60\\
adi & 1.01 & 1.00 & 1.01 & 1.00 & 1.01 & 1.00\\
fdtd-2d & 0.98 & 1.00 & 0.89 & 0.93 & 1.09 & 1.00\\
jacobi-1d-imper & 1.00 & 1.00 & 0.86 & 0.93 & 1.18 & 0.90\\
jacobi-2d-imper & 1.02 & 1.00 & 0.92 & 0.87 & 1.02 & 0.94\\
seidel-2d & 0.99 & 0.99 & 1.00 & 0.99 & 1.00 & 1.00\\

  \end{tabular}

\end{center} 

\subsection{x86-64 Intel Xeon Gold 6138}
High-performance, highly out-of-order, server-class Intel\textsuperscript{\textregistered} Xeon\textsuperscript{\textregistered} Gold 6138 CPU, running Debian GNU\discretionary{/}{}{/}Linux~10; gcc~8.3.0-6.

\begin{center}
  \begin{tabular}{l|rr|rr|rr}
    \multicolumn{1}{c|}{Benchmark} &
    \multicolumn{4}{c|}{\compcert} &
    \multicolumn{2}{c}{\gcc}\\
    &
    \multicolumn{2}{c|}{no unroll} &
    \multicolumn{2}{c|}{unroll+rotate} & \\
    & CSE3 & SSA
    & CSE3 & SSA
    & -O1 & -O2\\         
    \hline
    glpk & 0.73 & 0.89 & 0.88 & 0.91 & 0.61 & 0.76\\
picosat & 1.00 & 1.15 & 0.75 & 1.06 & 0.78 & 0.69\\
genann4 & 1.02 & 1.05 & 0.89 & 1.00 & 0.72 & 0.49\\
jpeg-6b & 0.96 & 1.00 & 1.04 & 0.88 & 0.82 & 0.91\\
ocaml & 0.90 & 0.83 & 0.91 & 0.99 & 0.89 & 0.78\\
\hline
correlation & 0.99 & 0.99 & 0.51 & 0.99 & 0.98 & 0.50\\
covariance & 0.99 & 1.00 & 0.51 & 0.98 & 0.96 & 0.50\\
2mm & 0.95 & 0.94 & 0.90 & 0.93 & 0.92 & 0.90\\
3mm & 0.97 & 0.95 & 0.91 & 0.94 & 0.93 & 0.92\\
atax & 0.99 & 1.00 & 0.60 & 1.13 & 0.92 & 0.53\\
bicg & 1.03 & 1.03 & 1.03 & 1.03 & 1.03 & 1.03\\
cholesky & 0.98 & 0.96 & 0.94 & 0.95 & 0.92 & 0.93\\
doitgen & 0.98 & 0.95 & 1.00 & 0.96 & 0.90 & 0.42\\
gemm & 0.94 & 1.24 & 0.98 & 0.92 & 0.99 & 1.01\\
gemver & 0.98 & 1.00 & 0.77 & 0.97 & 0.90 & 0.70\\
gesummv & 1.01 & 1.01 & 1.00 & 1.01 & 1.01 & 1.01\\
mvt & 0.99 & 0.97 & 0.71 & 0.95 & 0.93 & 0.66\\
symm & 0.94 & 0.99 & 0.95 & 0.94 & 0.94 & 0.93\\
syr2k & 1.00 & 1.00 & 1.00 & 1.00 & 0.99 & 0.61\\
syrk & 1.00 & 1.00 & 0.45 & 0.99 & 0.99 & 0.44\\
trisolv & 1.04 & 1.03 & 0.50 & 1.05 & 1.02 & 0.49\\
trmm & 1.00 & 1.01 & 0.46 & 0.99 & 0.98 & 0.45\\
durbin & 0.99 & 1.00 & 0.98 & 1.00 & 0.97 & 0.86\\
dynprog & 0.90 & 0.77 & 0.86 & 0.80 & 0.59 & 0.30\\
gramschmidt & 1.00 & 1.00 & 0.99 & 0.98 & 0.99 & 0.98\\
lu & 0.88 & 1.30 & 0.52 & 0.52 & 0.47 & 0.47\\
ludcmp & 0.89 & 0.76 & 0.74 & 0.81 & 0.63 & 0.63\\
floyd-warshall & 0.95 & 0.81 & 0.81 & 0.77 & 0.58 & 0.42\\
reg\_detect & 0.89 & 0.94 & 0.77 & 0.50 & 0.50 & 0.29\\
adi & 0.99 & 0.99 & 1.01 & 0.99 & 0.94 & 0.92\\
fdtd-2d & 1.02 & 0.92 & 0.75 & 0.77 & 0.67 & 0.66\\
jacobi-1d-imper & 0.79 & 0.90 & 0.65 & 0.70 & 0.29 & 0.58\\
jacobi-2d-imper & 1.20 & 1.00 & 1.02 & 0.96 & 0.64 & 0.62\\
seidel-2d & 1.00 & 1.00 & 1.01 & 1.01 & 1.01 & 1.02\\

  \end{tabular}

\end{center} 
 
\subsection{Risc-V ``Rocket''}
Risc-V 64~bit Rocket-core, inside a LowRisc~0.6 system-on-chip emulated on an Artix-7 FPGA on a Nexys-A7 board, running Debian testing;
  gcc 9.3.0.
  This core was chosen because it is similar to future embedded cores.
  
We experimented unexpected crashes (segmentation violations inside the standard library and system tools) with this system, so performance must be taken with a grain of salt.

\begin{center}
  \begin{tabular}{l|rr|rr|rr}
    \multicolumn{1}{c|}{Benchmark} &
    \multicolumn{4}{c|}{\compcert} &
    \multicolumn{2}{c}{\gcc}\\
    &
    \multicolumn{2}{c|}{no unroll} &
    \multicolumn{2}{c|}{unroll+rotate} & \\
    & CSE3 & SSA
    & CSE3 & SSA
    & -O1 & -O2\\         
    \hline
    glpk & 1.02 & 0.99 & 0.97 & 1.07 & 0.86 & 0.86\\
picosat & 1.12 & 1.08 & 0.98 & 0.98 & 0.87 & 0.82\\
genann4 & 1.04 & 1.03 & 0.97 & 0.99 & 0.98 & 0.79\\
jpeg-6b & 0.99 & 0.99 & 0.99 & 1.00 & 0.89 & 0.77\\
ocaml & 0.98 & 0.98 & 1.47 & 1.00 & 1.04 & 1.46\\
\hline
correlation & 0.88 & 0.87 & 0.88 & 0.88 & 0.78 & 0.75\\
covariance & 0.88 & 0.86 & 0.89 & 0.86 & 0.78 & 0.77\\
2mm & 0.92 & 0.90 & 0.91 & 0.91 & 0.83 & 0.77\\
3mm & 0.91 & 0.91 & 0.92 & 0.92 & 0.83 & 0.81\\
atax & 0.89 & 0.91 & 0.78 & 0.73 & 0.67 & 0.65\\
bicg & 0.93 & 0.94 & 0.93 & 0.92 & 0.80 & 0.76\\
cholesky & 0.89 & 0.85 & 0.70 & 1.80 & 0.64 & 0.67\\
doitgen & 0.82 & 0.77 & 0.87 & 0.77 & 0.58 & 0.53\\
gemm & 0.93 & 0.99 & 1.02 & 0.94 & 0.89 & 0.83\\
gemver & 0.94 & 0.93 & 0.88 & 0.89 & 0.81 & 0.73\\
gesummv & 0.95 & 0.96 & 0.97 & 0.95 & 0.84 & 0.80\\
mvt & 0.94 & 0.94 & 0.84 & 0.88 & 0.74 & 0.70\\
symm & 1.03 & 0.95 & 0.95 & 0.97 & 0.91 & 0.78\\
syr2k & 0.88 & 0.84 & 0.85 & 0.83 & 0.86 & 0.77\\
syrk & 1.00 & 0.98 & 0.89 & 0.89 & 0.74 & 0.68\\
trisolv & 0.94 & 0.94 & 0.74 & 0.72 & 0.67 & 0.67\\
trmm & 0.93 & 0.88 & 0.84 & 0.90 & 0.72 & 0.68\\
durbin & 0.92 & 1.03 & 0.92 & 0.91 & 0.81 & 0.76\\
dynprog & 0.77 & 0.72 & 0.71 & 0.71 & 0.44 & 0.27\\
gramschmidt & 0.91 & 0.92 & 0.92 & 0.92 & 0.90 & 0.69\\
lu & 0.96 & 1.01 & 1.13 & 0.94 & 0.74 & 0.71\\
ludcmp & 0.94 & 0.75 & 0.71 & 0.73 & 0.56 & 0.56\\
floyd-warshall & 0.95 & 0.89 & 0.93 & 0.91 & 0.69 & 0.65\\
reg\_detect & 0.86 & 0.76 & 0.57 & 0.44 & 0.24 & 0.20\\
adi & 0.96 & 0.96 & 0.97 & 0.95 & 0.91 & 0.87\\
fdtd-2d & 0.92 & 0.93 & 0.91 & 0.94 & 0.76 & 0.70\\
jacobi-1d-imper & 1.03 & 1.01 & 0.98 & 0.96 & 0.75 & 0.72\\
jacobi-2d-imper & 0.88 & 0.90 & 0.93 & 0.85 & 0.66 & 0.66\\
seidel-2d & 0.94 & 0.94 & 0.95 & 0.94 & 0.85 & 0.82\\

  \end{tabular}

\end{center}

\section{Trusted computing base}\label{sec:TCB}
The point of {\compcert} is to convey extremely strong assurance that the semantics of the assembly code matches that of the source code through theorems verified inside the {\coq} proof assistant.
If one trusts {\coq}, and more precisely the small proof checker inside {\coq}, then one can trust {\compcert}.
Yet, there are ways to use {\coq}, especially when dealing with extraction to {\ocaml} code and linking with external libraries, that can lead to undesirable additions to the trusted computing base.
Let us examine this issue in more detail.

\subsection{{\compcert}'s trusted computing base}
Vanilla {\compcert}'s trusted computing base consists of
\begin{enumerate}
\item {\coq}'s metatheory (e.g. the Calculus of Inductive Constructions is strongly normalizing);
\item a few axioms that have shown to be compatible with this metatheory (e.g. functional extensionality);
\item {\coq}'s implementation;
\item {\coq}'s extraction mechanism and {\ocaml}'s compiler and runtime system;
\item the extraction of some datatypes (pairs, Booleans\dots) into the corresponding {\ocaml} native datatypes;
\item\label{item:functional_ocaml} the functional character of the {\ocaml} functions called from {\coq};
\item the formal semantics of the first formal language ({\compcert} C);
\item option parsing and filename handling (in {\ocaml});
\item the frontend, which turns regular C into {\compcert} C and optionally deals with some constructs (bitfields, passing and returning structures to functions, variable length arguments\dots) through trusted {\ocaml} code;
\item the ``assembly expansion'' pass, trusted {\ocaml} code that expands certain pseudo-instructions into actual assembly code, including: stack allocation, stack deallocation, memory copy;
\item the axiomatization of these pseudo-instructions (e.g., the registers they may clobber);
\item the formal semantics of a formal assembly language;
\item the assembly language printer;
\item the compatibility of the application binary interface used by {\compcert} with that of the compiler used to compile other libraries on the system, including the standard library;
\item the assembler and linker.
\end{enumerate}

Each of these items is a possible unsoundness hazard, but the chances widely differ.
Among recently detected bugs in vanilla {\compcert} were one rarely used instruction being printed to assembly with incorrect instruction order, and two pseudo-assembly instructions being incorrectly axiomatized (scratch registers were clobbered but this was not reflected in the semantics).

In our opinion, the main suspects for possible bugs are: ABI compatibility, assembly printout (including tricky system-specific aspects), axiomatization and expansion of pseudo-assembly instructions, rather than, say, the implementations of {\coq} and {\ocaml}.
Indeed, it seems unlikely that there would be a bug in {\ocaml} that was not triggered by the many extant {\ocaml} applications, but that would trigger specifically when executing {\compcert} in a way that would not make {\compcert} crash or produce aberrant results, but instead silently produce wrong assembly code that would still be accepted by the assembler and linker.

\subsection{Analysis of our development}
In common to vanilla {\compcert}, we do not use logical axioms about the behavior of {\ocaml} code: that is, we never state axioms of the form ``this external {\ocaml} code returns a value that satisfies this property'' (e.g. we do not assume that abstract interpretation algorithms truly compute invariants; if we need such properties, we prove them).

Point~\ref{item:functional_ocaml} (the functional character of {\ocaml} code) applies throughout {\compcert}, including vanilla versions, which use numerous {\ocaml} functions;
it also applies to some of our extensions.
{\coq} is a purely functional programming language, thus when {\ocaml} functions are called from {\coq} it is assumed that they behave purely functionally from an external point of view: if a function $f$ is called twice on the same parameter $x$, then it returns the same value~$f(x)$.
This is not guaranteed in general in {\ocaml}, since a function may use impure operations and use persistent storage across calls.
There could be a proof where $f(x)$ and $f(x')$ appear, arising from two different calls, then the case where $x=x'$ is examined, and $f(x) \neq f(x')$ is dismissed as absurd, whereas this case is reachable in the extracted code.
This is not in general considered to be a serious issue: one is unlikely to distinguish such an ``absurd case'' by accident.
It is possible to work around this issue by wrapping the external {\ocaml} code in a nondeterministic monad, but this makes all programming and proofs considerably heavier, and anyway one would need to rewrite most of {\compcert} in monadic style to follow this idea.

The only place where we really add to {\compcert}'s trusted computing base is the hashed set library, in two ways:
\begin{enumerate}
\item\label{item:hash_consing} we assume a tiny bit of {\ocaml} code calling {\ocaml}'s weak hash tables and pointer equality is correct;
\item\label{item:weak_hash} we assume {\ocaml}'s weak hash tables behave correctly (but we can do without it, see below).
\end{enumerate}

Regarding point~\ref{item:weak_hash}: weak hash tables are used in both the {\ocaml} compiler and in {\coq}, thus if there are unsoundness issues they may already manifest themselves elsewhere in the trusted computing base.

There remains point~\ref{item:hash_consing}: the trusted correctness of the (very short) hash-consing code called from the custom constructor.
Could we do without it?
Hash-consing guarantees:
\begin{enumerate}
\item\label{item:requested_node} that the node returned by the hash-consing constructor is truly the requested node;
\item\label{item:structural_equality} that structural equality implies pointer equality (the converse properties always holds).
\end{enumerate}
Regarding point~\ref{item:requested_node}, our node equality test checks that the requested node and the node provided by hash-consing have identical contents (through pointer equalities);
if we did not trust the hash table we could run the equality test on its output and throw an exception if the nodes do not match.

Point~\ref{item:structural_equality} allows us to define a set equality operator as structural equality (by induction on the trees), then extract it as pointer equality.
Yet, we do not actually need a fully correct set equality operator. What we need in our proofs is that if set equality is deemed to hold, then the sets should be equal, which boils down to ``if the pointers to two sets are identical, then the sets are equal'', which is uncontroversial.
In no place we need to establish that two sets are not equal.

Would it be a problem if the weak hash table failed to retrieve a node already in the system and thus allow creating two different yet structurally equal sets? The possible consequences, neither of which a soundness hazard, are:
\begin{myitemize}
\item unnecessary recursion in set operations, where equality (never inequality) triggers shortcuts;
\item unnecessary fixed point iterations, where a fixed point is not detected, possibly leading to the maximal number of iterations being exceeded and a ``failed static analysis'' error being returned when compiling.
\end{myitemize}

\citet{six:hal-02185883}, when building the scheduling validator for the KV3, used another approach for their hash-consing:
  a special constructor function is used, but no assumption is made about its soundness (the returned term is checked);
  and physical equality is modeled as a nondeterministic function, such that when it returns true there is equality.
  Their approach has a smaller trusted computing base, but it loses the hash-consing axiom (which we do not really need) that when two terms are structurally equal, they are also at the same address in memory.
  It however is considerably heavier, due to the use of a nondeterminism monad.
  In addition, we intended our hashed set library to be usable independently of {\compcert}, and for other uses it is nice to have a true equality test with guaranteed equality/inequality answer, rather than a partial test.

\end{document}